\documentclass[journal=ancac3,manuscript=letter]{achemso}

\usepackage[version=3]{mhchem} 

\author{Cristian Cirac\`i}
\email{cristian.ciraci@iit.it}
\affiliation[IIT]
{Center for Biomolecular Nanotechnologies, Istituto Italiano di Tecnologia, Via Barsanti, 73010 Arnesano, Italy}
\author{Ferran Vidal-Codina}
\affiliation[MIT]
{Department of Aeronautics and Astronautics, Massachusetts Institute of Technology, Cambridge, MA 02139, USA}
\author{Daehan Yoo}
\affiliation[UMN]
{Department of Electrical and Computer Engineering, University of Minnesota, Minneapolis, MN 55455, USA}
\author{Jaime Peraire}
\affiliation[MIT]
{Department of Aeronautics and Astronautics, Massachusetts Institute of Technology, Cambridge, MA 02139, USA}
\author{Sang-Hyun Oh}
\affiliation[UMN]
{Department of Electrical and Computer Engineering, University of Minnesota, Minneapolis, MN 55455, USA}
\author{David R. Smith}
\affiliation[Duke]
{Center for Metamaterial and Integrated Plasmonics, Department of Electrical and Computer Engineering, Pratt School of Engineering, Duke University, Durham, NC 27708, USA}

\title[]{Impact of surface roughness in nanogap plasmonic systems}


\begin{document}
\begin{abstract}
Recent results have shown unprecedented control over separation distances between two metallic elements hundreds of nanometers in size, underlying the effects of free-electron nonlocal response also at mid-infrared wavelengths. 
Most of metallic systems however, still suffer from some degree of inhomogeneity due to fabrication-induced surface roughness.
Nanoscale roughness in such systems might hinder the understanding of the role of microscopic interactions.
Here we investigate the effect of surface roughness in coaxial nanoapertures resonating at mid-infrared frequencies.
We show that although random roughness shifts the resonances in an unpredictable way, the impact of nonlocal effects can still be clearly observed.
Roughness-induced perturbation on the peak resonance of the system shows a strong correlation with the effective gap size of the individual samples. 
Fluctuations due to fabrication imperfections then can be suppressed by performing measurements on structure ensembles in which averaging over a large number of samples   provides  a  precise  measure  of  the  ideal  system’s  optical  properties.
\end{abstract}


Plasmonics allows the confinement of light well below the diffraction limit by greatly enhancing the electric field in the vicinity of metal surfaces.
In the last decade, continuous developments in nanofabrication techniques have made it possible to control the separation distance between two metallic elements to a precision of a fraction of a nanometer \cite{Hill:2010ez,Moreau:2012uba,Duan:2012do,Chen:2013hq,Stewart:2016by,Chikkaraddy:6ja,Readman:2019cq}.
Such systems, generally referred to as \textit{nanogap plasmonic structures}, can squeeze light down to deep sub-wavelength volumes, allowing for the optical radiation to probe sub-atomic interactions\cite{Halas:2011cz, FernandezDominguez:2012eg, Esteban:2012gj, Wiener:2013gl, Barbry:2015iw, Zhu:2016fn, Carnegie:2018db,Yang:2019baa,Baumberg:2019iw}.
Most of metallic systems however, still suffer from some degree of inhomogeneity due to nanoscale surface roughness \cite{Wang:2006km,RodriguezFernandez:2009er,Trugler:2011ct}, which results in deviations of the optical properties with respect to ideally smooth systems \cite{Huang:2010hl,HyukPark:2012ik}.
Recent publications have reported on the important role of surface roughness on the far- and near-field as well as nonlinear optical properties of nanoparticles \cite{Martin:2002bn,Tinguely:2011gm,Lumdee:2015fn,Ning:2017bpa}.

In an experiment published in 2012 \cite{Ciraci:2012fp}, it was shown that the resonance of a film-coupled nanoparticle system undergoes a shift that cannot be explained by a simple \textit{local} constitutive relation between the electric field $\bf E$ and the polarization $\bf P$ of a metal: a more complex, \textit{nonlocal} \cite{Raza:2015ef}, relation accounting for electron-electron interactions had to be considered in order to predict the observed shifts.
This experiment triggered a debate in the community, with some reporting a classical behavior down to sub-nanometer gaps\cite{Doyle:2017hu}, other suggesting the \textit{anomalous} shift was simply due to the presence of roughness on the metallic film surface \cite{Hajisalem:2014gd, Hajisalem:2014cr}, although similar shifts were observed in ultra-smooth gold films obtained using template-stripping techniques \cite{Ciraci:2014jv}.
In general, however, observations beyond the classical response were reported in many independent works \cite{Kern:2012dz, Savage:2012by, Scholl:2013ge, Raza:2013gn,Cha:2014hd,Tan:2014hqa,Zhu:2014fza,Campos:2019et}. 
More recently, measurements on coaxial nanoapertures in the mid-infrared  \cite{Yoo:2019ec} have shown an even more consistent ($\sim 1~\mu$m) shift in the resonance compared to calculations performed within the local response approximation.
This work has reinvigorated the debate on whether the cause of the shift is due to sub-atomic electron-electron nonlocal interactions or simply due to a nonlocal correlation introduced by nanoscale surface roughness \cite{Kretschmann:1979fg}.

In this letter, we perform numerical calculations taking into account random surface roughness.
In particular, we investigate the coaxial nanoaperture system  and  show that the effect of including roughness is to randomly shift the peak resonance towards higher or lower frequencies.
The impact of this randomness on an ensemble of systems is to broaden the resonance without however affecting the resonance center of mass.


\begin{figure}
    \begin{center}
        \includegraphics[width=0.5\linewidth]{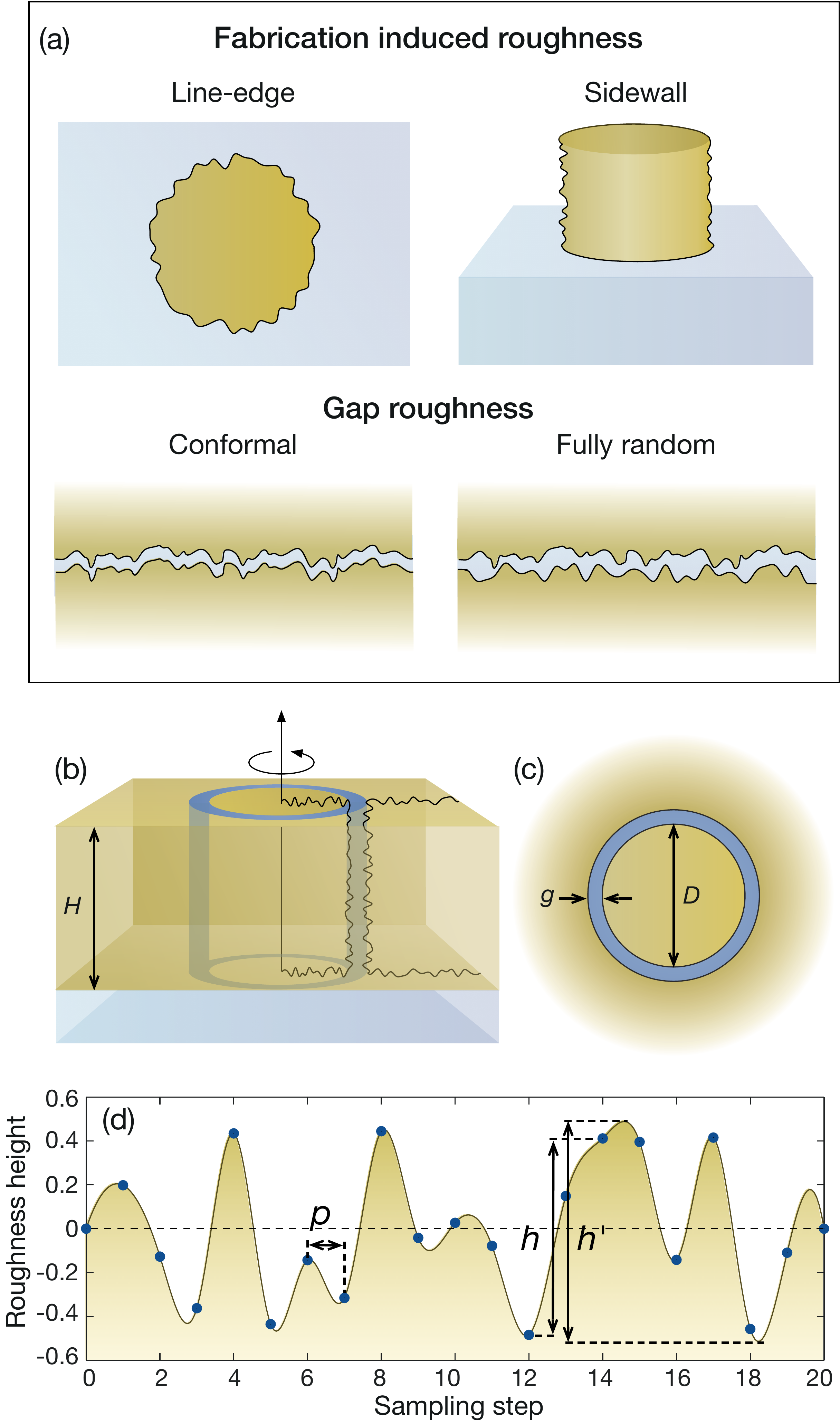}
    \end{center}
    \caption{ \small The coaxial nanoaperture surface roughness. In (a) a  summary of the different types of roughness involved in the fabrication process. (b) shows the coaxial nanogap geometry obtained by revolving a roughened cross-section; the resulting top view is depicted in (c). (d) Random roughness is created using a sampling distance $p$ and a maximum nominal height $h$ (or effective height $h'$). } 
    \label{fig1}
\end{figure}

Fabrication of coaxial nanoapertures involves four steps \cite{Yoo:2019ec}: first,  gold nanopillars are patterned via electron-beam lithography; the pillars are then coated by gap-filling insulator (Al$_2$O$_3$) using atomic layer deposition (ALD); the resulting structure is covered with gold by sputtering; finally, ion milling is used to planarize the top surface and expose the Al$_2$O$_3$-filled nanocoaxial apertures.
The lithography step introduces two different kind of irregularities: a line-edge roughness (seen from the top) that prevents the apertures from being perfectly circular, and a sidewall roughness.
Because of the nature of the ALD process, the gap is conformal to the sidewall irregularities.
However, a fully random source for the gap roughness cannot be excluded, so that in general the gap is affected by some combination of both types of irregularities.
For clarity we have summarized all of the kinds of roughnesses in Fig. \ref{fig1}a.

Numerically modeling random nanoscale surface roughness can be challenging for two reasons.
On one hand, it requires extended computational resources, since for example one cannot use periodic conditions and would need to discretize much more finely the computational space in order to resolve the roughness details.
On the other hand,  it is not straightforward creating randomness along an arbitrary surface.
In order to overcome these issues, we consider isolated structures, i.e. a single annular aperture in an infinitely extended metallic film, and we solve Maxwell's equations  using the so called \textit{2.5D technique} \cite{Ciraci:2013wi,Jurga:2017kn}.
This technique consists in expanding all the fields in cylindrical harmonics by exploiting the axis symmetry of the geometry. 
In this way one needs to solve just few two-dimensional problems while still maintaining the three-dimensionality of the original one.
The drawback of this approach is that roughness can be created only in the cross-sectional plane, that is, the structure will still be smooth along the azimuthal direction as shown in Fig. \ref{fig1}b-c.
A more general implementation could be in principle obtained by employing more complex numerical schemes that allow curvilinear elements\cite{Schmitt:2018im,Li:2017ci,VidalCodina:2018bh}. Such schemes, however, go beyond the scope of this work.
In this letter, we neglect line-edge roughness and consider the apertures to be perfectly circular.
Moreover, we analyze irregular structures with perfectly conformal gaps and fully random gap roughness separately.

Although roughness can be readily measured on planar thin films \cite{Nagpal:2009jr} using an atomic force microscope, it is not straightforward to do the same for sidewalls of gold pillars.
However, because in both cases roughness has an analogous origin we assume a roughness amplitude of the same order.  
In order to describe surface roughness we use two parameters: the roughness maximum height, $h'$, and the sampling step or average period of the roughness, $p$, as shown in Fig. \ref{fig1}d. 
Throughout the manuscript, we consider  $p=7.5$ nm.

\begin{figure}
    \begin{center}
        \includegraphics[width=0.5\linewidth]{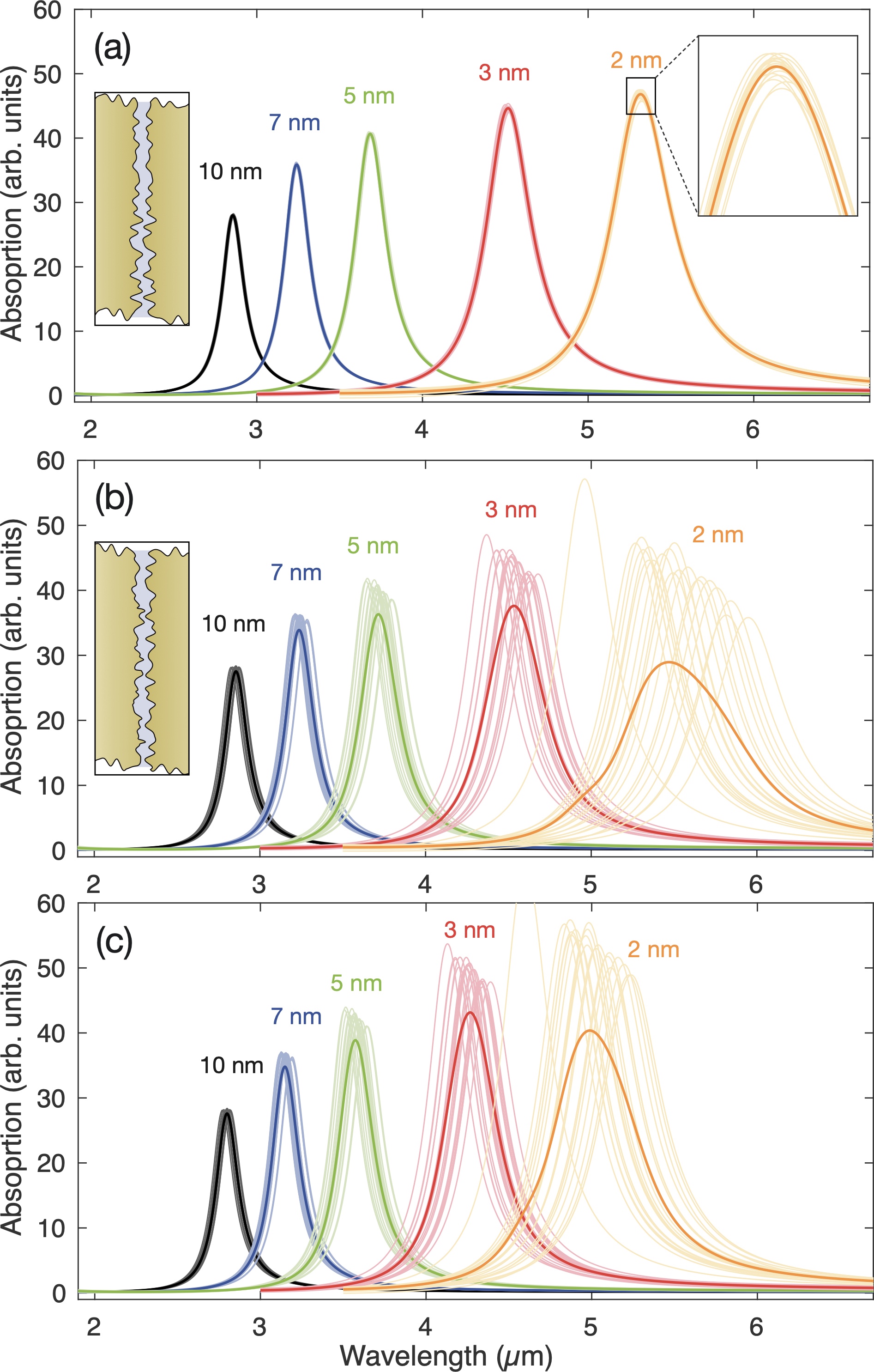}
    \end{center}
    \caption{\small Absorption spectra of a single nanoring aperture for different nominal gap sizes. The light curves refers to different random roughened geometries. The thick curve is obtained by averaging all different spectra for each gap size. In (a) a conformal gap roughness, as sketched in the inset, with $h\simeq1.8$ nm ($h'=2.5$ nm) is considered. In (b) a fully random roughness, as sketched in the inset, with $h=1.4$ nm ($h'\simeq2.0$ nm) has been considered. In (c) nonlocal response effects are included for the same numerical samples as in (b).} 
    \label{fig2}
\end{figure}

Our 2.5D approach greatly simplifies the creation of a randomly rough geometry.
It is in fact possible to draw the geometry cross-section by randomly displacing each point of the curve by a distance $|d|<=h/2$, following steps of $p$, where $h$ is the nominal maximum roughness height.
The result is a set of points that are interpolated using a \textit{spline} in order to smooth the curve and avoid numerical artifacts.
Note that by using spline interpolation we may get displacements between the nominal and effective maximum height, that is $h/2<|d|<=h'/2$.
An example of the result of this process is depicted in Fig. \ref{fig1}d.
We have implemented the 2.5D solver in a commercially available finite-element method software, Comsol Multiphysics. The geometry is created using a Matlab script 
in order to produce a set of points that are then interpolated directly in Comsol.

We consider coaxial nanoapertures drawn on an infinitely extended gold film of thickness $H=150$ nm, characterized by an internal diameter $D=250$ nm and a nominal gap $g$ varying from 2 to 10 nm, as depicted in Fig. \ref{fig1}b-c. 
The structure is excited by a plane wave impinging at normal incidence through an infinite sapphire substrate. The apertures are filled with Al$_2$O$_3$.
The permittivity of sapphire is obtained using the Sellmeier's formula with parameters as in Ref. \citenum{Yoo:2019ec}, while the dispersive dielectric constant of Al$_2$O$_3$ is extracted from the experimental measurements \cite{Kischkat:2012hz}.
Gold local permittivity is approximated by a Drude formula with $\hbar\omega_{\rm p}=8.45$ eV and $\hbar\gamma=0.047$ eV.
For completeness, we have also calculated the response of system when the electron pressure is taken into account, that is, when the metal response is nonlocal \cite{Raza:2015ef,VidalCodina:2018bh,Shen:2017ie}.
This is taken into account adding a correction of the form $\beta^2\nabla\nabla\cdot{\bf P}$ to the Drude model \cite{Raza:2015ef}, where $\beta$ is proportional to the Fermi velocity. Here we take $\beta\simeq1.3\times10^6$ m/s.
In order to characterize the optical response of the systems, we compute the absorption efficiency as $q_{\rm abs}=(W_{\rm abs}-W_{\rm film})/(\sigma I_0 )$, where $W_{\rm abs}$ is the power dissipated in the entire structure, $W_{\rm film}$ is the power dissipated by a continuous film (i.e. without aperture), $I_0$ is the input intensity and $\sigma=\pi\left [(D/2+g)^2-(D/2)^2\right ]$ is the geometrical cross-section of the aperture.
This definition removes all dependencies on the film extension. 

Using the techniques described above we have generated 20 different numerical \textit{samples} for each nominal gap size, for both types of gap roughness, conformal with $h=1.8$ nm ($h'\simeq2.5$) and fully random with $h=1.4$ nm ($h'\simeq2$). 
In Fig. \ref{fig2}, we show a set of absorption spectra around the zero-$th$-order mode resonance \cite{Yoo:2019ec,Yoo:2016iy,VidalCodina:2018bh}, as well as the mean curve obtained by averaging the results from all the random samples.
As expected, the resonance of the system undergoes a shift towards longer wavelengths as the average gap size is reduced.
It is interesting to note the difference between conformal (Fig. \ref{fig2}a) and random (Fig. \ref{fig2}b) gap roughness.
While in the first case (Fig. \ref{fig2}a)  the impact of roughness is minimal, producing small oscillations of the peak resonance even for the smallest gap, in the second case (Fig. \ref{fig2}b), the impact of roughness is substantial and variations of the peak resonance increase as the gap size shrinks.
It is clear from these simulations that the effect of the roughness is to randomly shift (towards higher or lower frequencies) the resonance of the system.
An analogous behavior can be observed for other resonances (i.e. the first order Fabry-P\'{e}rot mode resonance, not reported here).
In Fig. \ref{fig2}c, we report calculations for the same numerical samples of Fig. \ref{fig2}b when the nonlocal response of gold free electrons is considered.
As expected the average resonance results in an increasing blue-shift compared to the local case, as the gap is reduced.
More interestingly however is the fact that for each nominal gap size the spectra are more tightly packed compared to the local case.
This is especially visible for smaller gaps. 
The effect of the electron pressure is to spread the electron accumulation charge induced by an external field inward with respect to the metal surface. 
This can in some cases---for example when roughness produces sub-nanometer asperities---even mitigate the detrimental impact of surface roughness on the propagation length of surface plasmons \cite{Wiener:2012fd}.
Analogously, in this case the effect of the electron pressure is to smooth out some roughness features so that the global effect is a reduced random oscillation of the peak resonance caused by the randomness of the roughness.

\begin{figure*}
    \begin{center}
        \includegraphics[width=0.9\linewidth]{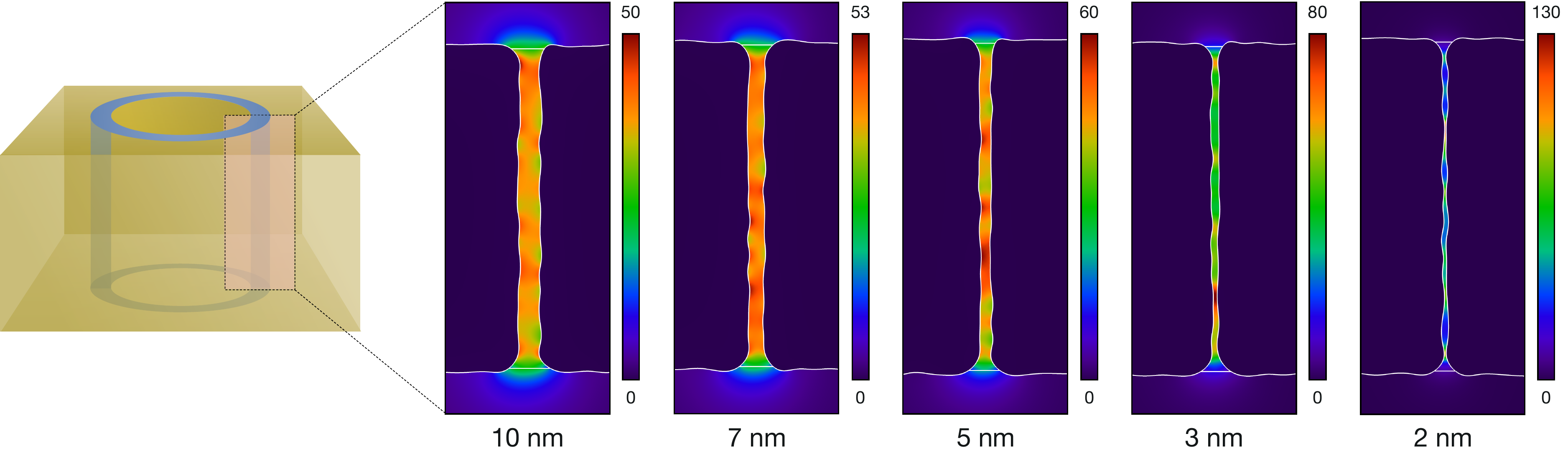}
    \end{center}
    \caption{\small Field enhancement ($|{\bf E}|/E_0$) maps for gap sizes $g = 10, 7, 5, 3, 2$ nm for structures with random surface roughness with $h=1.4$ nm ($h'\simeq2.0$ nm). The maps are taken in correspondence of the peak resonance for each structure.} 
    \label{fig3}
\end{figure*}

The impact of roughness on the near-field is shown in Fig. \ref{fig3}, where we show the electric field enhancement,  $|{\bf E}|/E_0$, at resonance, propagating through the aperture obtained for a typical sample for each gap size.
The resonance corresponds to an effective epsilon-near-zero mode, which is characterized by a constant phase, and hence norm, of the field along its propagation through the gap \cite{Yoo:2016iy}. 
For large gaps ($>5$ nm) this is still the case, while for smaller gaps ($< 3$ nm) the randomness of the structure creates local hot spots where the field is strongly enhanced.

\begin{figure}[H]
    \begin{center}
        \includegraphics[width=0.65\linewidth]{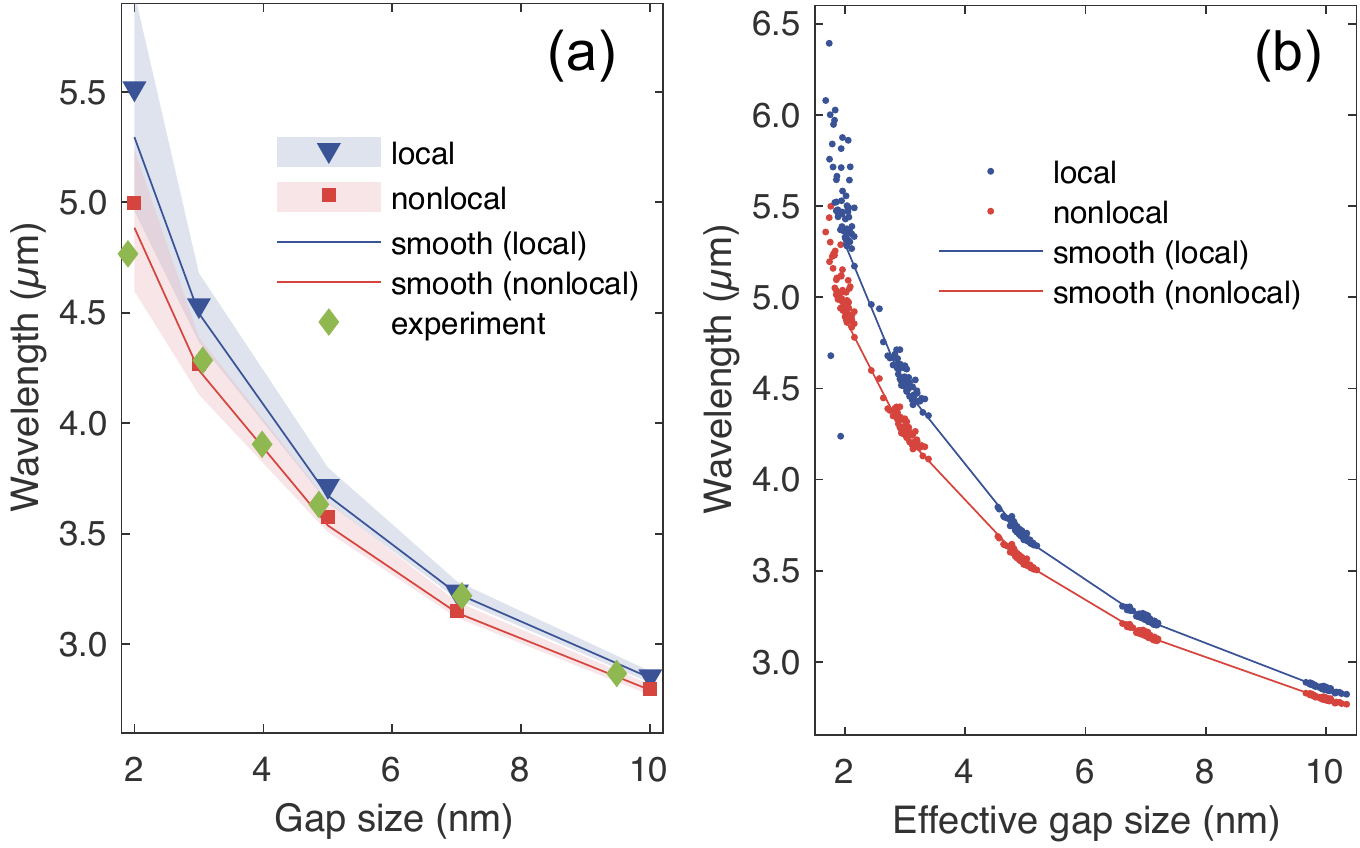}
    \end{center}
    \caption{\small  The resonant wavelength is tracked for the different models as a function of gap size.  In (a) the shaded areas represents the maximum deviation introduced by the purely random roughness. Nonlocality (in red) always blue-shifts resonances relatively to local response calculations (in blue); the continuous lines refer to a perfectly smooth structure; experimental data from  Ref. \citenum{Yoo:2019ec} are also reported. In (b) the resonance wavelength of individual structures is plotted against the effective gap size; smooth structures curves are shown for reference.} 
    \label{fig4}
\end{figure}

To understand the global impact of roughness on different gap sizes, in Fig. \ref{fig4}a we show the epsilon-near-zero mode resonance as a function of the nominal gap size $g$.
The shaded regions represent the maximum deviation introduced by the fully random roughness.
As already seen in Fig. \ref{fig2}, the smaller the gap the larger the variance.
The mean value (indicated by triangles and squares) is mostly unchanged with respect to the perfectly smooth structure for all gaps except the smallest.
In this case, the averaged resonance results are slightly shifted toward longer wavelengths compared to the smooth structure. 
Fig. \ref{fig4}a shows more clearly that the impact of electron pressure on coaxial nanoapertures is to slightly reduce (with respect to local calculations) the shift caused by the reduction of the gap size. 
This is consistent with numerical results obtained for smooth structures \cite{VidalCodina:2018bh}.
Together with our numerical data, we report in Fig. \ref{fig4}a the experimental data measured in Ref. \citenum{Yoo:2019ec}.
Note that the experimental data remain within the roughness deviation of the nonlocal calculations, and are clearly outside the span of local results. 

It is interesting at this point to understand the nature of the shift induced by the gap roughness.
In order to do so, we have evaluated the effective gap size corresponding to 3 sets of samples obtained for  $h = 1, 1.4, 1.8$ nm ($h' \simeq1.4, 2.0, 2.5$ nm). The effective gap for each sample ---one sample corresponds to a realization of random roughness for a given nominal gap and $h$--- is evaluated by numerically integrating the gap surface on a vertical cross-section, see Fig. \ref{fig1}b, and dividing it by the gold film thickness $H$. The epsilon-near-zero mode resonance for each sample is plotted against the effective gap size in Fig. \ref{fig4}b, for both local and nonlocal cases.
It is surprising how well the single samples follow the trajectory of the smooth structure.
This clearly shows that the resonance shift due to roughness is mostly driven by the variation that the roughness induces in the effective gap.
This effect is of course absent in the case of conformal roughness, where the effective gap always corresponds to the nominal gap. 

In conclusion, we have analyzed the impact of nanoscale surface roughness on single coaxial nanoapertures characterized by gaps of few nanometers.
Although our approach was limited to roughness on the cross-sectional plane of the systems, we are confident similar results will hold in the general case.
It is very unlikely in fact that full three-dimensional surface roughness might lead to different conclusions.
We have shown that while the impact of conformal gap roughness is negligible, fully random roughness might strongly affect the resonance shift of individual samples.
Moreover, the roughness-induced perturbation on the peak resonance has a strong correlation with the effective gap size for each individual sample.
A large number of samples then can average out  the fluctuations due to fabrication imperfections and still provide a precise measure of the ideal system's optical properties.
Experimentally, this is naturally achieved by performing measurements on structure ensembles.

\begin{acknowledgement}
F.V.-C. and J.P. acknowledge support from the AFOSR Grant No. FA9550-15-1-0276 and FA9550-16-0214. D.Y. and S-H.O. acknowledge support from the National Science Foundation (NSF ECCS 1809723 and ECCS 1809240) and Sanford P. Bordeau Endowed Chair at the University of Minnesota. D.R.S. acknowledges funding from AFOSR (Grant No. FA9550-18-1-0187). 
\end{acknowledgement}

\section{Notes} 
The authors declare no competing financial interest.




\providecommand{\latin}[1]{#1}
\makeatletter
\providecommand{\doi}
  {\begingroup\let\do\@makeother\dospecials
  \catcode`\{=1 \catcode`\}=2 \doi@aux}
\providecommand{\doi@aux}[1]{\endgroup\texttt{#1}}
\makeatother
\providecommand*\mcitethebibliography{\thebibliography}
\csname @ifundefined\endcsname{endmcitethebibliography}
  {\let\endmcitethebibliography\endthebibliography}{}


\begin{tocentry}
    \begin{center}
        \includegraphics[width=\linewidth]{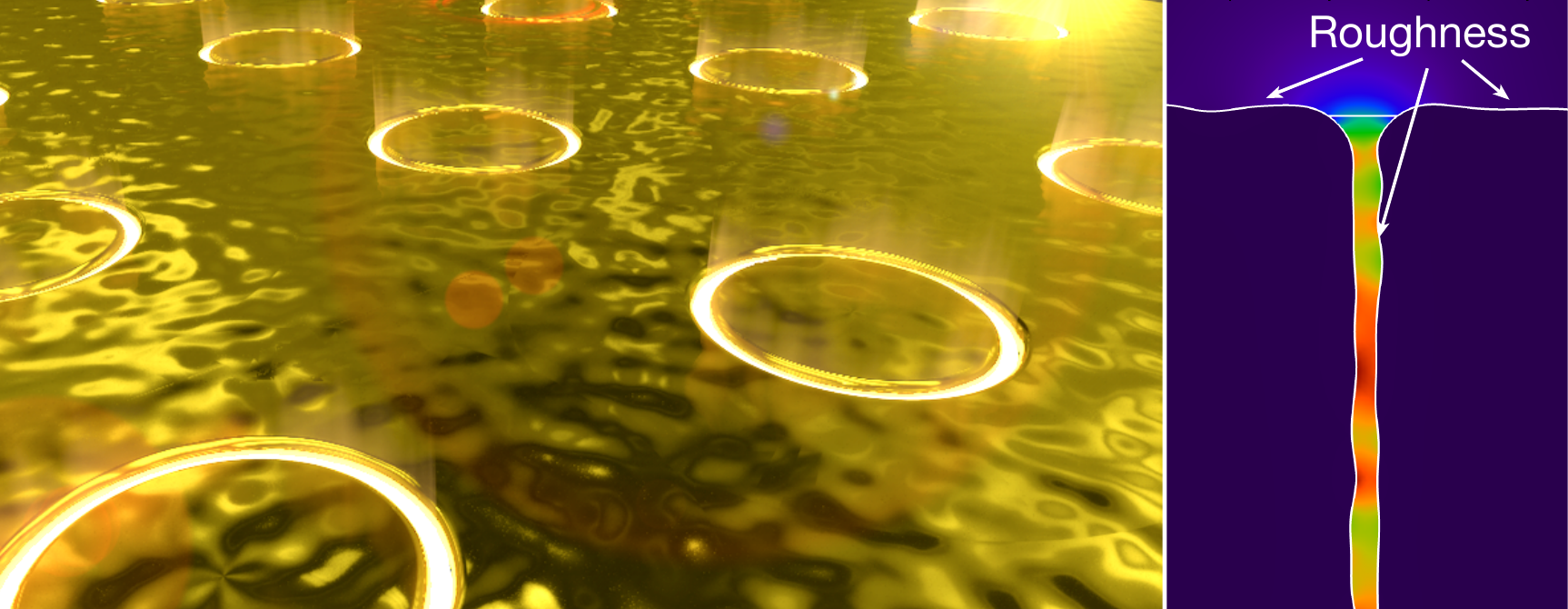}
    \end{center}
\end{tocentry}

\end{document}